\documentclass{nature_arxiv}

\usepackage[top=0.8in, bottom=1in, left=0.8in, right=0.8in]{geometry}

\usepackage{graphicx}
\usepackage{amsmath}
\usepackage{amssymb}

\newcommand{\dft}{DF2}
\newcommand{\dff}{DF4}

\newcommand{\kms}{km\,s$^{-1}$}

\title{A trail of dark matter-free galaxies from a bullet dwarf collision 
}

\author{\large Pieter van Dokkum$^{1}$,
Zili Shen$^{1}$,
Michael A.\ Keim$^{1}$,
Sebastian Trujillo-Gomez$^{2}$,
Shany Danieli$^{3}$,
Dhruba Dutta Chowdhury$^{1}$,
Roberto Abraham$^{4}$,
Charlie Conroy$^{5}$,
J.\ M.\ Diederik Kruijssen$^{2}$,
Daisuke Nagai$^{6}$,
Aaron Romanowsky$^{7,8}$
\vspace{8pt}}

\begin{document}

\maketitle

\begin{affiliations}
\small
 \item  Department of Astronomy, Yale University, New Haven, CT 06511, USA
\item Astronomisches Rechen-Institut, Zentrum f\"ur Astronomie der Universit\"at Heidelberg, M\"onchhofstra{\ss}e 12-14, D-69120 Heidelberg, Germany
 \item Department of Astrophysical Sciences, 4 Ivy Lane, Princeton University, Princeton, NJ 08544, USA
 \item Department of Astronomy \& Astrophysics, University of Toronto,  50 St.\ George Street, Toronto, ON M5S 3H4, Canada
\item Harvard-Smithsonian Center for Astrophysics, 60 Garden Street, Cambridge, MA, USA
\item Department of Physics, Yale University, New Haven, CT 06511, USA
\item  Department of Physics and Astronomy, San Jos\'e State University, San Jose, CA 95192, USA
\item University of California Observatories, 1156 High Street, Santa Cruz, CA 95064, USA
 
\end{affiliations}


\begin{abstract}
The ultra-diffuse galaxies \dft\ and \dff\ in the NGC1052 group share
several unusual properties: they both have large sizes,\cite{cohen:18} rich
populations of overluminous and large globular clusters,\cite{dokkum:18b,dokkum:19df4,dutta:19,dhruba:20,shen:21a} and very low velocity
dispersions indicating little or no dark matter.\cite{dokkum:18,dokkum:18gc98,danieli:19,emsellem:19} 
It has been suggested that these galaxies were formed in the aftermath of high velocity collisions
of  gas rich galaxies,\cite{silk:19,shin:20,lee:21}
events that resemble the collision that created the bullet cluster\cite{clowe:06} but on much smaller scales.
The gas separates from the dark matter in the collision and subsequent star formation leads to the formation
of one or more dark matter-free galaxies.\cite{shin:20}
Here we show that
the present-day line-of-sight distances and radial velocities of \dft\ and \dff\ are consistent with their
joint formation in the aftermath of a single bullet-dwarf
collision, around eight billion years ago. 
Moreover, we find that \dft\ and \dff\ are part of an apparent linear substructure of seven to eleven
large, low-luminosity objects. We propose that these all originated in the same event,
forming a trail of dark matter-free galaxies
that is roughly more than two megaparsecs long and angled $7^{\circ}\pm2^{\circ}$ from the line of sight.
We also tentatively identify the highly dark
matter-dominated remnants of the two progenitor galaxies that are expected\cite{silk:19} at the leading edges of the
trail.
\end{abstract}

\begin{figure*}[t]
  \begin{center}
  \includegraphics[width=1.00\linewidth]{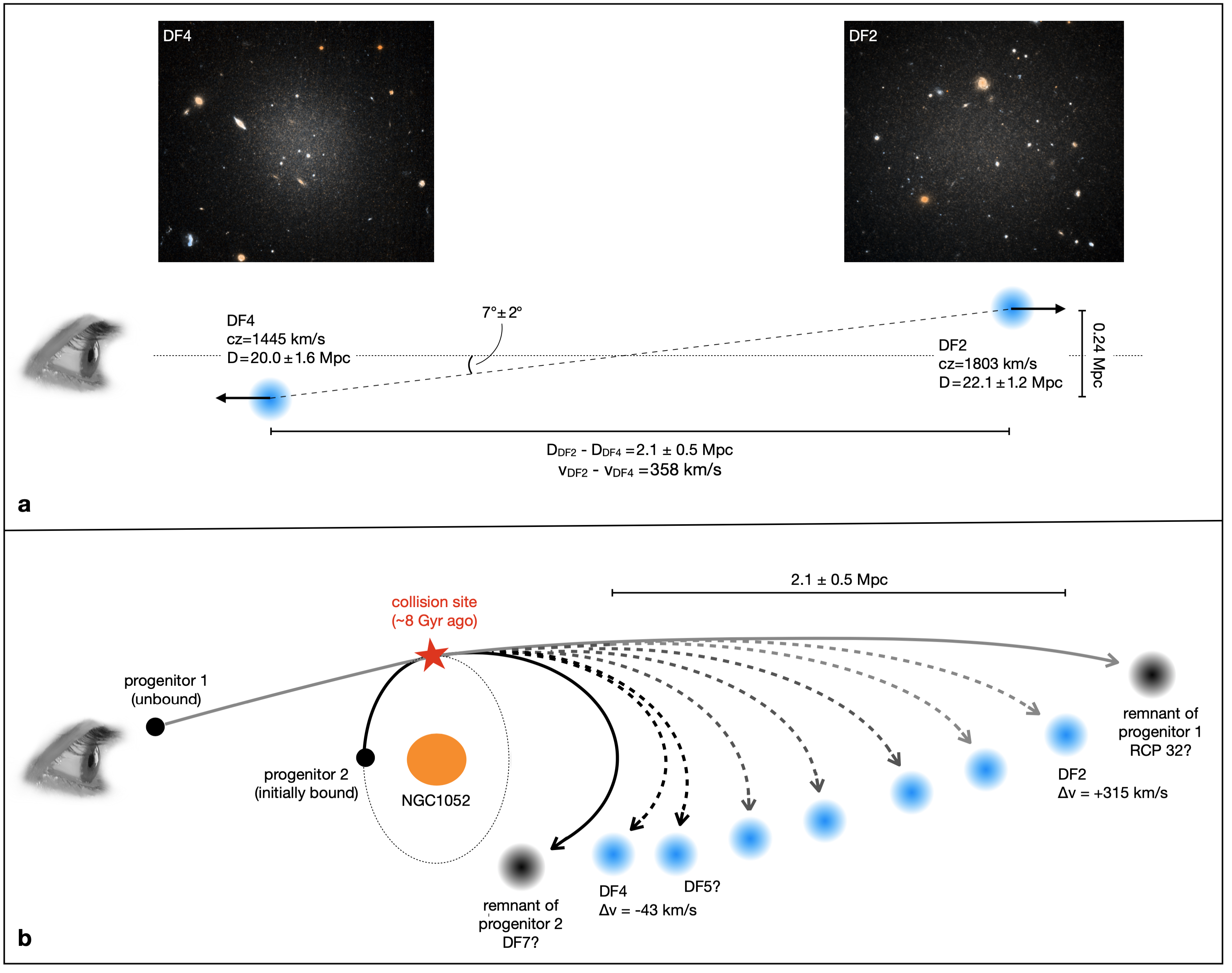}
  \end{center}
    \vspace{-0.3truecm}  
    \caption{\small \textbf{Geometry of the \dft\ and \dff\ system.} 
{\bf a.} The radial velocity difference between \dft\ and \dff\ is 358\,\kms\ and this large
velocity difference is accompanied by a large line-of-sight separation of $2.1\pm 0.5$\,Mpc.\cite{danieli:20,shen:21b} 
The geometry implies that the galaxies are moving away from each other. Tracing their positions back in time,
we infer that they were formed in a high-speed encounter $\geq  6$\,Gyr ago.
{\bf b.} Example of a collisional scenario involving NGC1052. Velocities are given with respect to that galaxy ($cz=1,488$\,\kms). An infalling
gas-rich galaxy on an unbound orbit collided with a satellite of NGC1052 about 8\,Gyr ago, leading to two dark remnants
(possibly RCP32 and DF7), DF2 and DF4, and three to seven other dark matter-free galaxies.
   }
   \label{geometry.fig}
    \vspace{-12pt}
\end{figure*}

We begin with the assumption that it is not a coincidence that  \dft\ and \dff\ have the same
set of otherwise-unique properties and that they were in close proximity to one another at the
time of their formation. With that assumption, collisional formation is implied by their present-day
radial velocities and three-dimensional locations. The geometry is shown in Fig.\ \ref{geometry.fig}a.
The relative radial velocity of the galaxies is high, $358$\,\kms, which is three times the velocity
dispersion of the NGC1052 group (about 115\,\kms). Furthermore,
while the two galaxies are separated by only 0.24\,Mpc in the plane of the sky\cite{dokkum:19df4} 
a differential tip of the red giant branch (TRGB) analysis has shown that they are $2.1\pm 0.5$\,Mpc apart along
the line of sight,\cite{danieli:20,shen:21b}  which is five times the virial radius
of NGC1052 (about  400\,kpc\cite{forbes:19,shen:21b}). 
In the context of a shared origin of both
galaxies, their radial velocities are consistent with their line-of-sight distances,
that is, the closest galaxy (\dff) is moving toward us (with respect to the mean velocity) and the farthest galaxy
(\dft) is moving away from us.
Tracing their line-of-sight positions back in time, we 
infer that they must have formed in a high-velocity encounter. The minimum
time since that encounter is about 6\,Gyr, for constant motion along the line of sight.

The situation at that point, a $\gtrsim 300$\,\kms\ encounter between
two galaxies in the gas-rich environment of a young group,
is a close match to the initial conditions of mini-bullet-cluster\cite{clowe:06}
scenarios that have previously been proposed for the formation of DF2 and DF4.\cite{silk:19,shin:20,lee:21}
In a near head-on collision between two gas-rich galaxies,
the collisional gas can be shocked and
separated from the collisionless dark matter and
pre-existing stars.\cite{silk:19,shin:20} The accompanying star formation favours
massive clumps in highly compressed gas,\cite{silk:19,lee:21} producing the unusual globular clusters
as well as the lack of dark matter. Galaxies that form this way are initially compact,\cite{lee:21} in
apparent conflict with the large half-light radii of DF2 and DF4. However, intense
feedback accompanying the formation of the globular clusters
is expected to increase
the sizes of the newly formed galaxies, an effect that is particularly efficient when the stars are not bound by
a dark matter halo.\cite{trujillogomez:21,trujillogomez:21b}
These previous studies focused on the formation of a single new galaxy at the
collision site, but we propose that both \dft\ and \dff\ 
were formed in a single `bullet dwarf' event.
An illustration of the
proposed scenario is shown in the Methods section.

We have sufficient
information to construct a plausible model for the geometry and timing of the collision. It is likely that the event
took place near the central elliptical galaxy, NGC1052, as it is roughly halfway between DF2 and DF4 in projection
and its deep potential well is conducive to high-speed interactions. It is also likely that at least one of the
progenitor galaxies was a satellite of NGC1052
as the probability of a collision of two unbound galaxies is extremely small.\cite{shin:20} 
We assume that the second progenitor was not bound to NGC1052, and either on first infall or
a satellite of another massive galaxy in the group.\cite{dokkum:1042}
DF4 has a velocity difference of $-43$\,\kms\  with respect to NGC1052, whereas the velocity of DF2 is $+315$\,\kms.
In Fig.\ \ref{geometry.fig}b we show a configuration for DF2, DF4, and NGC1052 that satisfies
these constraints.
Progenitor 1 arrived in the vicinity of NGC1052 with a high ($>300$\,\kms) relative velocity and
collided with progenitor 2, which was on a bound orbit. The collision produced DF2 and DF4, with DF2's velocity
and orbit similar to progenitor 1 and DF4's similar to progenitor 2. DF2 is currently unbound and about 3\,Mpc
behind the group, whereas DF4 remained bound and has begun falling back. This geometry is not unique
but it is similar to examples that have been explored in simulations.\cite{shin:20}
It is also consistent with the near-identical tidal distortions of the two galaxies:\cite{keim:21}
in the geometry of Fig.\ \ref{geometry.fig}b, DF2 and DF4 were at the same distance from NGC1052 when they were formed, and as neither galaxy
experienced a stronger tidal field afterwards, their morphologies have remained the same.
In this model, DF2 has travelled about 3\,Mpc since the collision with an average velocity with respect to NGC1052
that is probably higher than its present-day value of $315$\,\kms.
Assuming an average post-collisional velocity of  $\langle v \rangle \sim 350$\,\kms\ dates the collision to about 8\,Gyr ago, in excellent agreement with the 
ages of the globular clusters and the diffuse light in \dft\ 
($9\pm 2$\,Gyr, as measured from optical spectra\cite{dokkum:18b,fensch:19}).

\begin{figure*}[htb]
  \begin{center}
  \includegraphics[width=1.0\linewidth]{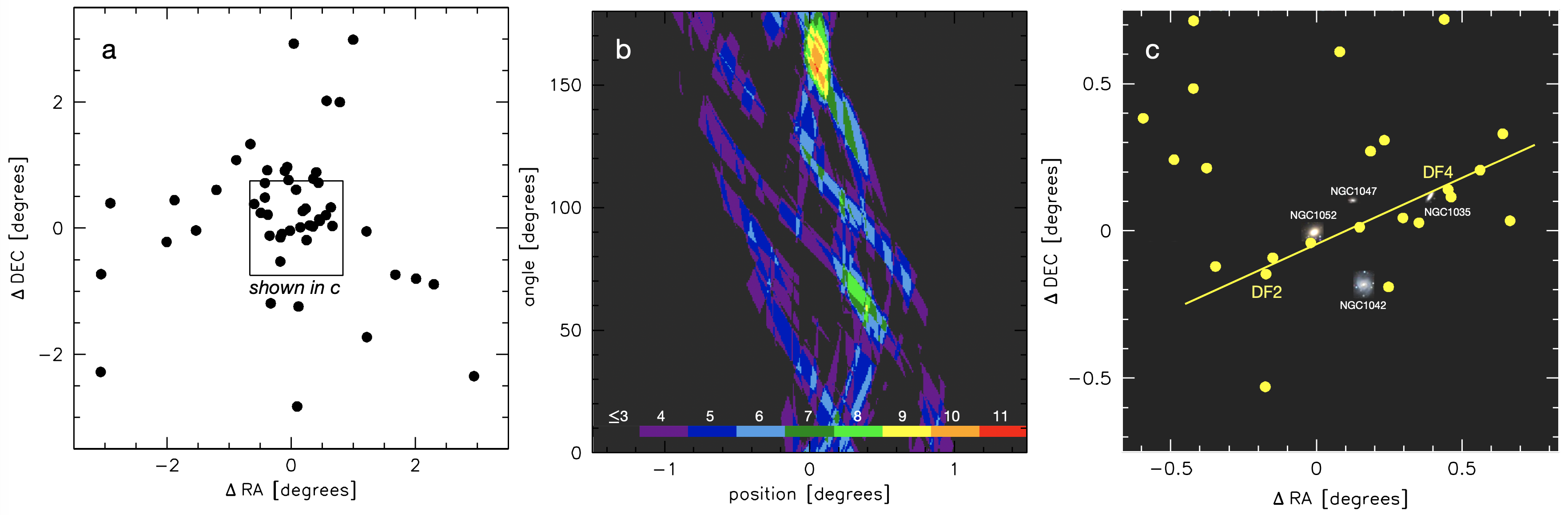}
  \end{center}
    \vspace{-0.5truecm}  
    \caption{\small \textbf{A linear feature in the spatial distribution of faint galaxies in the NGC1052 field.} 
{\bf a.} Distribution of galaxies with $g>16.5$ (black circles) from a recent compilation
of low surface brightness objects in the NGC1052 field.\cite{roman:21} 
The positions are with respect to the coordinates of NGC1052, the central bright elliptical galaxy
in the group. {\bf b.} Hough transform\cite{dudahart:72} of the spatial distribution. The
peak corresponds to a line that has 11 galaxies located within $\pm 30$\,kpc. The significance
of the feature, as determined from randomized realizations of the data, is $97$\,\%.  {\bf c.} Zoom of
{\bf a}, shown with the four brightest members of the NGC1052 group. Yellow circles correspond to the positions of galaxies in the box shown in panel {\bf a}. The orientation and offset
of the line corresponds to the location of the peak of the
Hough transform. Both \dft\ and \dff\ are part of the linear feature.
   }
   \label{spatial.fig}
    \vspace{-12pt}
\end{figure*}

\begin{figure*}[htb]
  \begin{center}
  \includegraphics[width=1.00\linewidth]{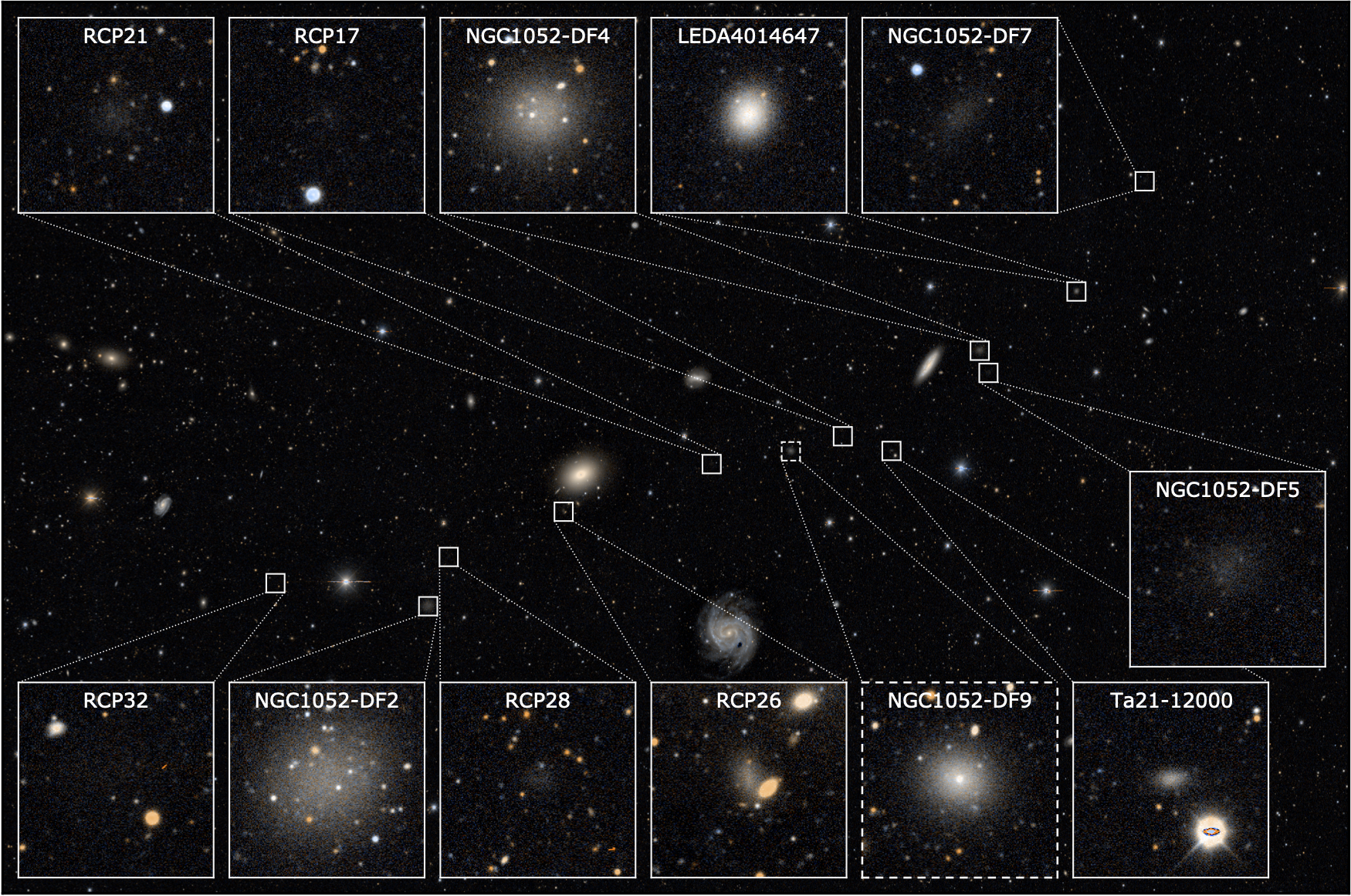}
  \end{center}
    \vspace{-0.5truecm}  
    \caption{\small \textbf{Galaxies on the \dft\,--\,\dff\ axis.} 
Legacy survey image of the central region of the NGC1052 group, highlighting the 11 galaxies that
are part of the trail according to the Hough transform. Several ($2\pm 2$) of these are expected to be chance projections.
LEDA4014647 is a candidate for an interloper (that is, an unrelated group member) given its brightness and
relative compactness.
Its radial velocity was listed as $1680\pm 60$ in earlier SDSS releases (DR3) but was later erroneously revised to
a $z=0.7$ QSO (DR16). Judging from morphology alone, RCP26 and Ta21-12000 may also be
chance projections. Besides \dft\ and \dff, RCP32, DF5, and DF7 all satisfy or nearly satisfy the UDG
criteria. RCP32 and DF7 are
candidates for the two dark matter-dominated remnants that have been predicted to precede dark
matter-deficient galaxies along the post-collision trajectory.\cite{silk:19} We also highlight DF9
(SDSS\,J024007.01$-$081344.4),  a galaxy with a bright star cluster that falls on the trail
but is not part of the objectively-selected sample.\cite{roman:21}
   }
   \label{cutouts.fig}
    \vspace{-12pt}
\end{figure*}


We further investigate the possible joint formation of DF2 and DF4 by examining
the spatial distribution of galaxies along the \dft\,--\,\dff\ axis.
Given the complex gas distribution during and after
the event, it
may be that more than two dark matter-deficient objects
were formed in the wake of the collision.\cite{shin:20} 
Furthermore, the bullet-dwarf scenario predicts the existence of two dark matter-dominated objects that are the remnants of the progenitor galaxies. These should precede DF2, DF4 and any other dark matter-deficient galaxies along their path, as they have the highest velocities relative to the barycentre.\cite{silk:19}

The spatial distribution of galaxies with magnitude $g>16.5$ in a recently compiled
catalog of the NGC1052 field\cite{roman:21}  is shown in Fig.\ \ref{spatial.fig}a. 
An objective search for linear features in this distribution is performed using a discrete implementation
of the Hough transform.\cite{dudahart:72} 
The transform is shown in Fig.\ \ref{spatial.fig}b.
There is a clear peak with 11 galaxies on a line, corresponding to the relation \begin{equation}
\Delta{\rm DEC}=0.45 \Delta{\rm RA} - 2.8,
\end{equation}
in units of arcminutes north and west of NGC1052. This relation is shown by the line in Fig.\ \ref{spatial.fig}c.
Both \dft\ and \dff\ are in the sample of 11 galaxies.
The probability that the peak arose by chance is 3\,\%, and the probability that the peak arose by chance
and that both \dft\ and \dff\ are part of it is 0.6\,\% (see Methods).
Before turning to the properties of the galaxies in the trail, we note that the Hough transform
provides post hoc validation of our initial assumption that \dft\ and \dff\ are related to each other.

\begin{figure*}[htb]
  \begin{center}
  \includegraphics[width=0.95\linewidth]{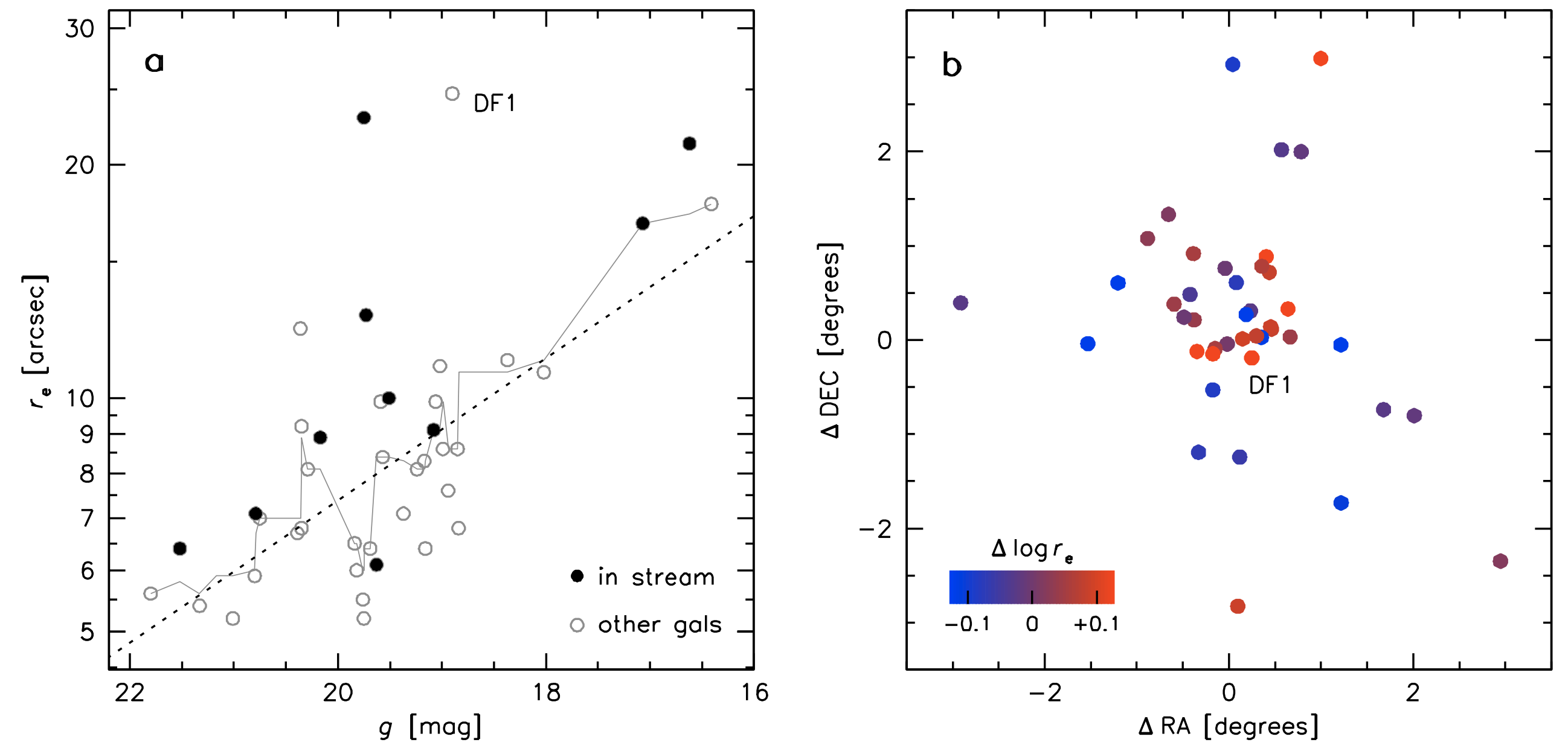}
  \end{center}
    \vspace{-0.5truecm}  
    \caption{\small \textbf{The sizes of galaxies on the \dft\,--\,\dff\ axis.} 
{\bf a.} The apparent size -- apparent magnitude relation for low surface brightness objects in NGC1052, using
recent measurements  from Legacy data.\cite{roman:21} Galaxies that are part of the trail are labeled with
solid symbols. The thin solid line is a running median, with $N=7$. The broken line is a least squares fit to the running
median and has the form $\log r_{\rm e} = -0.09 (g-20) + 0.87$ kpc. 
Galaxies in the trail are typically
larger than other galaxies of the same magnitude. {\bf b.} 
Distribution of galaxies color-coded by their location with respect to the least squares fit. The largest
galaxies in the group are preferentially located along the DF2\,--\,DF4 axis. 
   }
   \label{sizes.fig}
    \vspace{-12pt}
\end{figure*}

Images of the 11 galaxies that are part of the trail are shown in Fig.\
\ref{cutouts.fig}. The average galaxy
density in the central $R<30'$ implies that $2\pm 2$ of the 11 are chance projections, and we infer that
there are 7\,--\,11 galaxies in the structure. Besides \dft\ and \dff\ other galaxies in the trail
are also uncommonly large for their luminosity. The relation between size and apparent magnitude for
faint galaxies in the NGC1052 group is shown in Fig.\ \ref{sizes.fig}a.   After subtracting a simple linear least squares fit to the running median (broken
line), we find that galaxies in the trail are on average 26\,\% larger than the rest of the sample. The Wilcoxon
probability that the trail galaxies and the rest of the sample are drawn from the same size
distribution is $<1$\,\%. The spatial distribution of galaxies color-coded by their (magnitude-dependent)
relative size is shown in Fig.\ \ref{sizes.fig}b.
The unusual prevalence of large, low surface brightness galaxies in the central regions of the NGC1052 group
had been noticed previously;\cite{cohen:18,roman:21} here we propose that the bullet dwarf event was responsible for it.

We highlight three large and unusual galaxies that are part of
the sample of eleven.
DF5 looks very similar to DF2 and DF4 but has a much lower surface brightness.\cite{cohen:18,keim:21}
It is so close in projection to DF4 (only $1.6'$ away) that both galaxies were observed in the same HST/ACS pointing.\cite{cohen:18,danieli:19} 
In the bullet dwarf scenario this puzzling object is readily explained as another dark matter-free fragment that formed in the aftermath of the collision.
RCP32 and DF7 are the furthest away from the
center of the structure, and are ``ahead'' of \dft\ and \dff\ respectively. RCP32 is the faintest ultra-diffuse galaxy (UDG)
that has so far been discovered in integrated light and may have a globular cluster
system.\cite{roman:21}  DF7 is an elongated UDG that was previously observed with HST \cite{cohen:18}
(see Methods section). 
We tentatively identify RCP32 and DF7 as the candidate remnants of the original, pre-collision galaxies.
In the geometry
of Fig.\ \ref{geometry.fig}b RCP32 could be the remnant
of progenitor 1, the gas-rich object that was not bound to NGC1052, and the brighter galaxy DF7 could be the remnant of
progenitor 2, the galaxy that was a NGC1052 satellite at the time of collision.
In this context the elongation
can be understood as the effect of the strong tidal forces at the time of collision.\cite{rich:12}

The scenario that is proposed here makes predictions for the properties of the collision products, and
further observations can test and refine this explanation. As the formation of \dft\ and \dff\ was
triggered by a single event the ages of the globular clusters of \dff, which have
not yet been measured, should be identical to those of the clusters in \dft\ ($9\pm 2$\,Gyr.\cite{dokkum:18b,fensch:19}).
A stringent and model-independent test is to directly compare the (averaged) spectra of clusters
in the two galaxies; any clear differences could falsify our model, particularly differences in age-sensitive
spectral features. Turning to other galaxies on the trail,  their kinematics are predicted to be consistent with
baryon-only models --- except RCP32 and DF7, which could show evidence for an unusually {\em low}
baryon fraction. Furthermore,
the radial velocities of trail galaxies
should follow the approximate relation $cz \approx 1700 - 10\Delta{\rm RA}$\,\kms, with $\Delta{\rm RA}$ 
defined as in Eq.\,1. Similarly, their line-of-sight distances are predicted to be $D \approx 21.5 - 0.06\Delta{\rm RA}$\,Mpc, for
TRGB measurements that are calibrated to $D=22.1$\,Mpc for DF2.\cite{shen:21b}
We note that these relations likely have considerable scatter and are not expected to be linear for most geometries.
Also,  0 -- 4 galaxies of the 11 are expected to be interlopers (see Fig.\ \ref{cutouts.fig}), and 
some galaxies that seem to be off the trail may
in fact be part of it due to the foreshortening (see Fig.\ 1a).
This may be the case for DF1, a very large and diffuse galaxy (marked in Fig.\ \ref{sizes.fig}) that is only $14'$ off the
\dft\,--\,\dff\ axis and is elongated toward the center of the trail.

Bullet dwarf collisions hold the potential to constrain the self-interaction cross section
of dark matter.
Modeling of the bullet cluster has provided an upper limit,\cite{randall:08}  but
as self-interacting
dark matter  was introduced to explain the ``cored'' dark matter density profiles of low
mass galaxies\cite{spergel:00} it is important to measure the cross section on small
scales.\cite{tulin:18}
Quantitative constraints will
likely require more than a single example of a bullet
dwarf. Encouragingly, these events
are probably more common than the collision that produced the bullet cluster;\cite{bouillot:15}
a search for plausible \dft\,/\,\dff\  progenitors in the Illustris TNG cosmological simulation\cite{tng}
produced 248 head-on high velocity collisions in a $100^3$\,Mpc$^3$
volume,\cite{shin:20} corresponding to $\sim 8$ within $D<20$\,Mpc.


\vspace{1cm}

\bibliography{master}

\newcommand{\noop}[1]{}
\begin{thebibliography}{10}
\small
\expandafter\ifx\csname url\endcsname\relax
  \def\url#1{\texttt{#1}}\fi
\expandafter\ifx\csname urlprefix\endcsname\relax\def\urlprefix{URL }\fi
\providecommand{\bibinfo}[2]{#2}
\providecommand{\eprint}[2][]{\url{#2}}

\bibitem{cohen:18}
\bibinfo{author}{{Cohen}, Y.} \emph{et~al.}
\newblock \bibinfo{title}{{The Dragonfly Nearby Galaxies Survey. V. HST/ACS
  Observations of 23 Low Surface Brightness Objects in the Fields of NGC 1052,
  NGC 1084, M96, and NGC 4258}}.
\newblock \emph{\bibinfo{journal}{\apj}} \textbf{\bibinfo{volume}{868}},
  \bibinfo{pages}{96} (\bibinfo{year}{2018}).

\bibitem{dokkum:18b}
\bibinfo{author}{{van Dokkum}, P.} \emph{et~al.}
\newblock \bibinfo{title}{{An Enigmatic Population of Luminous Globular
  Clusters in a Galaxy Lacking Dark Matter}}.
\newblock \emph{\bibinfo{journal}{\apjl}} \textbf{\bibinfo{volume}{856}},
  \bibinfo{pages}{L30} (\bibinfo{year}{2018}).

\bibitem{dokkum:19df4}
\bibinfo{author}{{van Dokkum}, P.}, \bibinfo{author}{{Danieli}, S.},
  \bibinfo{author}{{Abraham}, R.}, \bibinfo{author}{{Conroy}, C.} \&
  \bibinfo{author}{{Romanowsky}, A.~J.}
\newblock \bibinfo{title}{{A Second Galaxy Missing Dark Matter in the NGC 1052
  Group}}.
\newblock \emph{\bibinfo{journal}{\apjl}} \textbf{\bibinfo{volume}{874}},
  \bibinfo{pages}{L5} (\bibinfo{year}{2019}).

\bibitem{dutta:19}
\bibinfo{author}{{Dutta Chowdhury}, D.}, \bibinfo{author}{{van den Bosch},
  F.~C.} \& \bibinfo{author}{{van Dokkum}, P.}
\newblock \bibinfo{title}{{On the Orbital Decay of Globular Clusters in NGC
  1052-DF2: Testing a Baryon-only Mass Model}}.
\newblock \emph{\bibinfo{journal}{\apj}} \textbf{\bibinfo{volume}{877}},
  \bibinfo{pages}{133} (\bibinfo{year}{2019}).

\bibitem{dhruba:20}
\bibinfo{author}{{Dutta Chowdhury}, D.}, \bibinfo{author}{{van den Bosch},
  F.~C.} \& \bibinfo{author}{{van Dokkum}, P.}
\newblock \bibinfo{title}{{On the Evolution of the Globular Cluster System in
  NGC 1052-DF2: Dynamical Friction, Globular-Globular Interactions, and
  Galactic Tides}}.
\newblock \emph{\bibinfo{journal}{\apj}} \textbf{\bibinfo{volume}{903}},
  \bibinfo{pages}{149} (\bibinfo{year}{2020}).

\bibitem{shen:21a}
\bibinfo{author}{{Shen}, Z.}, \bibinfo{author}{{van Dokkum}, P.} \&
  \bibinfo{author}{{Danieli}, S.}
\newblock \bibinfo{title}{{A Complex Luminosity Function for the Anomalous
  Globular Clusters in NGC 1052-DF2 and NGC 1052-DF4}}.
\newblock \emph{\bibinfo{journal}{\apj}} \textbf{\bibinfo{volume}{909}},
  \bibinfo{pages}{179} (\bibinfo{year}{2021}).

\bibitem{dokkum:18}
\bibinfo{author}{{van Dokkum}, P.} \emph{et~al.}
\newblock \bibinfo{title}{{A galaxy lacking dark matter}}.
\newblock \emph{\bibinfo{journal}{\nat}} \textbf{\bibinfo{volume}{555}},
  \bibinfo{pages}{629--632} (\bibinfo{year}{2018}).

\bibitem{dokkum:18gc98}
\bibinfo{author}{{van Dokkum}, P.} \emph{et~al.}
\newblock \bibinfo{title}{{A Revised Velocity for the Globular Cluster GC-98 in
  the Ultra Diffuse Galaxy NGC{\,}1052-DF2}}.
\newblock \emph{\bibinfo{journal}{Research Notes of the American Astronomical
  Society}} \textbf{\bibinfo{volume}{2}}, \bibinfo{pages}{54}
  (\bibinfo{year}{2018}).

\bibitem{danieli:19}
\bibinfo{author}{{Danieli}, S.}, \bibinfo{author}{{van Dokkum}, P.},
  \bibinfo{author}{{Conroy}, C.}, \bibinfo{author}{{Abraham}, R.} \&
  \bibinfo{author}{{Romanowsky}, A.~J.}
\newblock \bibinfo{title}{{Still Missing Dark Matter: KCWI High-resolution
  Stellar Kinematics of NGC1052-DF2}}.
\newblock \emph{\bibinfo{journal}{\apjl}} \textbf{\bibinfo{volume}{874}},
  \bibinfo{pages}{L12} (\bibinfo{year}{2019}).

\bibitem{emsellem:19}
\bibinfo{author}{{Emsellem}, E.} \emph{et~al.}
\newblock \bibinfo{title}{{The ultra-diffuse galaxy NGC 1052-DF2 with MUSE. I.
  Kinematics of the stellar body}}.
\newblock \emph{\bibinfo{journal}{\aap}} \textbf{\bibinfo{volume}{625}},
  \bibinfo{pages}{A76} (\bibinfo{year}{2019}).

\bibitem{silk:19}
\bibinfo{author}{{Silk}, J.}
\newblock \bibinfo{title}{{Ultra-diffuse galaxies without dark matter}}.
\newblock \emph{\bibinfo{journal}{\mnras}} \textbf{\bibinfo{volume}{488}},
  \bibinfo{pages}{L24--L28} (\bibinfo{year}{2019}).

\bibitem{shin:20}
\bibinfo{author}{{Shin}, E.-j.} \emph{et~al.}
\newblock \bibinfo{title}{{Dark Matter Deficient Galaxies Produced via
  High-velocity Galaxy Collisions in High-resolution Numerical Simulations}}.
\newblock \emph{\bibinfo{journal}{\apj}} \textbf{\bibinfo{volume}{899}},
  \bibinfo{pages}{25} (\bibinfo{year}{2020}).

\bibitem{lee:21}
\bibinfo{author}{{Lee}, J.}, \bibinfo{author}{{Shin}, E.-j.} \&
  \bibinfo{author}{{Kim}, J.-h.}
\newblock \bibinfo{title}{{Dark Matter Deficient Galaxies and Their Member Star
  Clusters Form Simultaneously during High-velocity Galaxy Collisions in 1.25
  pc Resolution Simulations}}.
\newblock \emph{\bibinfo{journal}{\apjl}} \textbf{\bibinfo{volume}{917}},
  \bibinfo{pages}{L15} (\bibinfo{year}{2021}).

\bibitem{clowe:06}
\bibinfo{author}{{Clowe}, D.} \emph{et~al.}
\newblock \bibinfo{title}{{A Direct Empirical Proof of the Existence of Dark
  Matter}}.
\newblock \emph{\bibinfo{journal}{\apjl}} \textbf{\bibinfo{volume}{648}},
  \bibinfo{pages}{L109--L113} (\bibinfo{year}{2006}).

\bibitem{danieli:20}
\bibinfo{author}{{Danieli}, S.} \emph{et~al.}
\newblock \bibinfo{title}{{A Tip of the Red Giant Branch Distance to the Dark
  Matter Deficient Galaxy NGC 1052-DF4 from Deep Hubble Space Telescope Data}}.
\newblock \emph{\bibinfo{journal}{\apjl}} \textbf{\bibinfo{volume}{895}},
  \bibinfo{pages}{L4} (\bibinfo{year}{2020}).

\bibitem{shen:21b}
\bibinfo{author}{{Shen}, Z.} \emph{et~al.}
\newblock \bibinfo{title}{{A Tip of the Red Giant Branch Distance of 22.1
  {\ensuremath{\pm}} 1.2 Mpc to the Dark Matter Deficient Galaxy NGC 1052-DF2
  from 40 Orbits of Hubble Space Telescope Imaging}}.
\newblock \emph{\bibinfo{journal}{\apjl}} \textbf{\bibinfo{volume}{914}},
  \bibinfo{pages}{L12} (\bibinfo{year}{2021}).

\bibitem{forbes:19}
\bibinfo{author}{{Forbes}, D.~A.}, \bibinfo{author}{{Alabi}, A.},
  \bibinfo{author}{{Brodie}, J.~P.} \& \bibinfo{author}{{Romanowsky}, A.~J.}
\newblock \bibinfo{title}{{Dark matter and no dark matter: on the halo mass of
  NGC 1052}}.
\newblock \emph{\bibinfo{journal}{\mnras}} \textbf{\bibinfo{volume}{489}},
  \bibinfo{pages}{3665--3669} (\bibinfo{year}{2019}).

\bibitem{trujillogomez:21}
\bibinfo{author}{{Trujillo-Gomez}, S.}, \bibinfo{author}{{Kruijssen},
  J.~M.~D.}, \bibinfo{author}{{Keller}, B.~W.} \&
  \bibinfo{author}{{Reina-Campos}, M.}
\newblock \bibinfo{title}{{Constraining the formation of NGC 1052-DF2 from its
  unusual globular cluster population}}.
\newblock \emph{\bibinfo{journal}{\mnras}} \textbf{\bibinfo{volume}{506}},
  \bibinfo{pages}{4841--4854} (\bibinfo{year}{2021}).

\bibitem{trujillogomez:21b}
\bibinfo{author}{{Trujillo-Gomez}, S.}, \bibinfo{author}{{Kruijssen}, J.~M.~D.}
  \& \bibinfo{author}{{Reina-Campos}, M.}
\newblock \bibinfo{title}{{The emergence of dark matter-deficient ultra-diffuse
  galaxies driven by scatter in the stellar mass-halo mass relation and
  feedback from globular clusters}}.
\newblock \emph{\bibinfo{journal}{MNRAS, in press}}
  \bibinfo{pages}{arXiv:2103.08610} (\bibinfo{year}{2021}).

\bibitem{dokkum:1042}
\bibinfo{author}{{van Dokkum}, P.}, \bibinfo{author}{{Danieli}, S.},
  \bibinfo{author}{{Romanowsky}, A.}, \bibinfo{author}{{Abraham}, R.} \&
  \bibinfo{author}{{Conroy}, C.}
\newblock \bibinfo{title}{{The Distance to NGC 1042 in the Context of its
  Proposed Association with the Dark Matter-deficient Galaxies NGC 1052-DF2 and
  NGC 1052-DF4}}.
\newblock \emph{\bibinfo{journal}{Research Notes of the American Astronomical
  Society}} \textbf{\bibinfo{volume}{3}}, \bibinfo{pages}{29}
  (\bibinfo{year}{2019}).

\bibitem{keim:21}
\bibinfo{author}{{Keim}, M.~A.} \emph{et~al.}
\newblock \bibinfo{title}{{Tidal Distortions in NGC1052-DF2 and NGC1052-DF4:
  Independent Evidence for a Lack of Dark Matter}}.
\newblock \emph{\bibinfo{journal}{arXiv e-prints}}
  \bibinfo{pages}{arXiv:2109.09778} (\bibinfo{year}{2021}).

\bibitem{fensch:19}
\bibinfo{author}{{Fensch}, J.} \emph{et~al.}
\newblock \bibinfo{title}{{The ultra-diffuse galaxy NGC 1052-DF2 with MUSE. II.
  The population of DF2: stars, clusters, and planetary nebulae}}.
\newblock \emph{\bibinfo{journal}{\aap}} \textbf{\bibinfo{volume}{625}},
  \bibinfo{pages}{A77} (\bibinfo{year}{2019}).

\bibitem{roman:21}
\bibinfo{author}{{Rom{\'a}n}, J.}, \bibinfo{author}{{Castilla}, A.} \&
  \bibinfo{author}{{Pascual-Granado}, J.}
\newblock \bibinfo{title}{{Discovery and analysis of low-surface-brightness
  galaxies in the environment of NGC 1052}}.
\newblock \emph{\bibinfo{journal}{\aap}} \textbf{\bibinfo{volume}{656}},
  \bibinfo{pages}{A44} (\bibinfo{year}{2021}).

\bibitem{dudahart:72}
\bibinfo{author}{{Duda}, R.~O.} \& \bibinfo{author}{{Hart}, P.~E.}
\newblock \bibinfo{title}{{Use of the Hough transformation to detect lines and
  curves in pictures}}.
\newblock \emph{\bibinfo{journal}{Communications of the ACM}}
  \textbf{\bibinfo{volume}{15}}, \bibinfo{pages}{11--15}
  (\bibinfo{year}{1972}).

\bibitem{rich:12}
\bibinfo{author}{{Rich}, R.~M.} \emph{et~al.}
\newblock \bibinfo{title}{{A tidally distorted dwarf galaxy near NGC 4449}}.
\newblock \emph{\bibinfo{journal}{\nat}} \textbf{\bibinfo{volume}{482}},
  \bibinfo{pages}{192--194} (\bibinfo{year}{2012}).

\bibitem{randall:08}
\bibinfo{author}{{Randall}, S.~W.}, \bibinfo{author}{{Markevitch}, M.},
  \bibinfo{author}{{Clowe}, D.}, \bibinfo{author}{{Gonzalez}, A.~H.} \&
  \bibinfo{author}{{Brada{\v{c}}}, M.}
\newblock \bibinfo{title}{{Constraints on the Self-Interaction Cross Section of
  Dark Matter from Numerical Simulations of the Merging Galaxy Cluster 1E
  0657-56}}.
\newblock \emph{\bibinfo{journal}{\apj}} \textbf{\bibinfo{volume}{679}},
  \bibinfo{pages}{1173--1180} (\bibinfo{year}{2008}).

\bibitem{spergel:00}
\bibinfo{author}{{Spergel}, D.~N.} \& \bibinfo{author}{{Steinhardt}, P.~J.}
\newblock \bibinfo{title}{{Observational Evidence for Self-Interacting Cold
  Dark Matter}}.
\newblock \emph{\bibinfo{journal}{\prl}} \textbf{\bibinfo{volume}{84}},
  \bibinfo{pages}{3760--3763} (\bibinfo{year}{2000}).

\bibitem{tulin:18}
\bibinfo{author}{{Tulin}, S.} \& \bibinfo{author}{{Yu}, H.-B.}
\newblock \bibinfo{title}{{Dark matter self-interactions and small scale
  structure}}.
\newblock \emph{\bibinfo{journal}{\physrep}} \textbf{\bibinfo{volume}{730}},
  \bibinfo{pages}{1--57} (\bibinfo{year}{2018}).

\bibitem{bouillot:15}
\bibinfo{author}{{Bouillot}, V.~R.}, \bibinfo{author}{{Alimi}, J.-M.},
  \bibinfo{author}{{Corasaniti}, P.-S.} \& \bibinfo{author}{{Rasera}, Y.}
\newblock \bibinfo{title}{{Probing dark energy models with extreme pairwise
  velocities of galaxy clusters from the DEUS-FUR simulations}}.
\newblock \emph{\bibinfo{journal}{\mnras}} \textbf{\bibinfo{volume}{450}},
  \bibinfo{pages}{145--159} (\bibinfo{year}{2015}).

\bibitem{tng}
\bibinfo{author}{{Pillepich}, A.} \emph{et~al.}
\newblock \bibinfo{title}{{Simulating galaxy formation with the IllustrisTNG
  model}}.
\newblock \emph{\bibinfo{journal}{\mnras}} \textbf{\bibinfo{volume}{473}},
  \bibinfo{pages}{4077--4106} (\bibinfo{year}{2018}).

\bibitem{decals}
\bibinfo{author}{{Dark Energy Survey Collaboration}}.
\newblock \bibinfo{title}{{The Dark Energy Survey: more than dark energy - an
  overview}}.
\newblock \emph{\bibinfo{journal}{\mnras}} \textbf{\bibinfo{volume}{460}},
  \bibinfo{pages}{1270--1299} (\bibinfo{year}{2016}).

\bibitem{galfit}
\bibinfo{author}{{Peng}, C.~Y.}, \bibinfo{author}{{Ho}, L.~C.},
  \bibinfo{author}{{Impey}, C.~D.} \& \bibinfo{author}{{Rix}, H.-W.}
\newblock \bibinfo{title}{{Detailed Structural Decomposition of Galaxy
  Images}}.
\newblock \emph{\bibinfo{journal}{\aj}} \textbf{\bibinfo{volume}{124}},
  \bibinfo{pages}{266--293} (\bibinfo{year}{2002}).

\bibitem{beers:90}
\bibinfo{author}{{Beers}, T.~C.}, \bibinfo{author}{{Flynn}, K.} \&
  \bibinfo{author}{{Gebhardt}, K.}
\newblock \bibinfo{title}{{Measures of location and scale for velocities in
  clusters of galaxies - A robust approach}}.
\newblock \emph{\bibinfo{journal}{\aj}} \textbf{\bibinfo{volume}{100}},
  \bibinfo{pages}{32--46} (\bibinfo{year}{1990}).

\bibitem{bailin:08}
\bibinfo{author}{{Bailin}, J.}, \bibinfo{author}{{Power}, C.},
  \bibinfo{author}{{Norberg}, P.}, \bibinfo{author}{{Zaritsky}, D.} \&
  \bibinfo{author}{{Gibson}, B.~K.}
\newblock \bibinfo{title}{{The anisotropic distribution of satellite
  galaxies}}.
\newblock \emph{\bibinfo{journal}{\mnras}} \textbf{\bibinfo{volume}{390}},
  \bibinfo{pages}{1133--1156} (\bibinfo{year}{2008}).

\bibitem{tempel:15}
\bibinfo{author}{{Tempel}, E.}, \bibinfo{author}{{Guo}, Q.},
  \bibinfo{author}{{Kipper}, R.} \& \bibinfo{author}{{Libeskind}, N.~I.}
\newblock \bibinfo{title}{{The alignment of satellite galaxies and cosmic
  filaments: observations and simulations}}.
\newblock \emph{\bibinfo{journal}{\mnras}} \textbf{\bibinfo{volume}{450}},
  \bibinfo{pages}{2727--2738} (\bibinfo{year}{2015}).

\bibitem{wang:08}
\bibinfo{author}{{Wang}, Y.} \emph{et~al.}
\newblock \bibinfo{title}{{Probing the intrinsic shape and alignment of dark
  matter haloes using SDSS galaxy groups}}.
\newblock \emph{\bibinfo{journal}{\mnras}} \textbf{\bibinfo{volume}{385}},
  \bibinfo{pages}{1511--1522} (\bibinfo{year}{2008}).

\bibitem{silverman:86}
\bibinfo{author}{{Silverman}, B.~W.}
\newblock \emph{\bibinfo{title}{{Density Estimation for Statistics and Data
  Analysis}}} (\bibinfo{publisher}{Chapman \& Hall, London},
  \bibinfo{year}{1986}).

\bibitem{trujillo:21}
\bibinfo{author}{{Trujillo}, I.} \emph{et~al.}
\newblock \bibinfo{title}{{Introducing the LBT Imaging of Galactic Halos and
  Tidal Structures (LIGHTS) survey. A preview of the low surface brightness
  Universe to be unveiled by LSST}}.
\newblock \emph{\bibinfo{journal}{\aap}} \textbf{\bibinfo{volume}{654}},
  \bibinfo{pages}{A40} (\bibinfo{year}{2021}).

\bibitem{martin:18}
\bibinfo{author}{{Martin}, N.~F.}, \bibinfo{author}{{Collins}, M.~L.~M.},
  \bibinfo{author}{{Longeard}, N.} \& \bibinfo{author}{{Tollerud}, E.}
\newblock \bibinfo{title}{{Current Velocity Data on Dwarf Galaxy NGC 1052-DF2
  do not Constrain it to Lack Dark Matter}}.
\newblock \emph{\bibinfo{journal}{\apjl}} \textbf{\bibinfo{volume}{859}},
  \bibinfo{pages}{L5} (\bibinfo{year}{2018}).

\bibitem{trujillo:19}
\bibinfo{author}{{Trujillo}, I.} \emph{et~al.}
\newblock \bibinfo{title}{{A distance of 13 Mpc resolves the claimed anomalies
  of the galaxy lacking dark matter}}.
\newblock \emph{\bibinfo{journal}{\mnras}} \textbf{\bibinfo{volume}{486}},
  \bibinfo{pages}{1192--1219} (\bibinfo{year}{2019}).

\bibitem{monelli:19}
\bibinfo{author}{{Monelli}, M.} \& \bibinfo{author}{{Trujillo}, I.}
\newblock \bibinfo{title}{{The TRGB Distance to the Second Galaxy
  {\textquotedblleft}Missing Dark Matter{\textquotedblright}: Evidence for Two
  Groups of Galaxies at 13.5 and 19 Mpc in the Line of Sight of NGC 1052}}.
\newblock \emph{\bibinfo{journal}{\apjl}} \textbf{\bibinfo{volume}{880}},
  \bibinfo{pages}{L11} (\bibinfo{year}{2019}).

\bibitem{dokkum:18c}
\bibinfo{author}{{van Dokkum}, P.} \emph{et~al.}
\newblock \bibinfo{title}{{A Revised Velocity for the Globular Cluster GC-98 in
  the Ultra Diffuse Galaxy NGC\,1052-DF2}}.
\newblock \emph{\bibinfo{journal}{Research Notes of the American Astronomical
  Society}} \textbf{\bibinfo{volume}{2}}, \bibinfo{pages}{54}
  (\bibinfo{year}{2018}).

\bibitem{ogiya:18}
\bibinfo{author}{{Ogiya}, G.}
\newblock \bibinfo{title}{{Tidal stripping as a possible origin of the ultra
  diffuse galaxy lacking dark matter}}.
\newblock \emph{\bibinfo{journal}{\mnras}} \textbf{\bibinfo{volume}{480}},
  \bibinfo{pages}{L106--L110} (\bibinfo{year}{2018}).

\bibitem{maccio:21}
\bibinfo{author}{{Macci{\`o}}, A.~V.} \emph{et~al.}
\newblock \bibinfo{title}{{Creating a galaxy lacking dark matter in a dark
  matter-dominated universe}}.
\newblock \emph{\bibinfo{journal}{\mnras}} \textbf{\bibinfo{volume}{501}},
  \bibinfo{pages}{693--700} (\bibinfo{year}{2021}).

\bibitem{jackson:21}
\bibinfo{author}{{Jackson}, R.~A.} \emph{et~al.}
\newblock \bibinfo{title}{{Dark matter-deficient dwarf galaxies form via tidal
  stripping of dark matter in interactions with massive companions}}.
\newblock \emph{\bibinfo{journal}{\mnras}} \textbf{\bibinfo{volume}{502}},
  \bibinfo{pages}{1785--1796} (\bibinfo{year}{2021}).

\bibitem{moreno:22}
\bibinfo{author}{{Moreno}, J.} \emph{et~al.}
\newblock \bibinfo{title}{{Galaxies lacking dark matter produced by close
  encounters in a cosmological simulation}}.
\newblock \emph{\bibinfo{journal}{Nature Astronomy}}  (\bibinfo{year}{2022}).

\bibitem{montes:20}
\bibinfo{author}{{Montes}, M.} \emph{et~al.}
\newblock \bibinfo{title}{{The Galaxy ``Missing Dark Matter'' NGC 1052-DF4 is
  Undergoing Tidal Disruption}}.
\newblock \emph{\bibinfo{journal}{\apj}} \textbf{\bibinfo{volume}{904}},
  \bibinfo{pages}{114} (\bibinfo{year}{2020}).

\bibitem{montes:21}
\bibinfo{author}{{Montes}, M.}, \bibinfo{author}{{Trujillo}, I.},
  \bibinfo{author}{{Infante-Sainz}, R.}, \bibinfo{author}{{Monelli}, M.} \&
  \bibinfo{author}{{Borlaff}, A.~S.}
\newblock \bibinfo{title}{{A Disk and No Signatures of Tidal Distortion in the
  Galaxy ``Lacking'' Dark Matter NGC 1052-DF2}}.
\newblock \emph{\bibinfo{journal}{\apj}} \textbf{\bibinfo{volume}{919}},
  \bibinfo{pages}{56} (\bibinfo{year}{2021}).

\end{thebibliography}

\renewcommand\thefigure{Extended Data Figure \arabic{figure}}
\setcounter{figure}{0}

\begin{methods}

\subsection{Illustration of the collision scenario.}
The proposed scenario for the formation of DF2, DF4, and the other trail galaxies is shown in Fig.\ \ref{cartoon.fig}.
As discussed in the main text, the scenario is a combination of the original idea that a bullet dwarf collision might have formed DF2 and/or DF4;\cite{silk:19}
the results from subsequent hydrodynamical simulations, showing that multiple dark matter-free clumps can form
in such a collision\cite{shin:20} and that the formation of massive star clusters is indeed promoted;\cite{lee:21} and
the independent finding that feedback from massive cluster formation in these conditions leads to a rapid expansion
of the galaxies.\cite{trujillogomez:21}

\begin{figure}[htb]
  \begin{center}
  \includegraphics[width=1.0\linewidth]{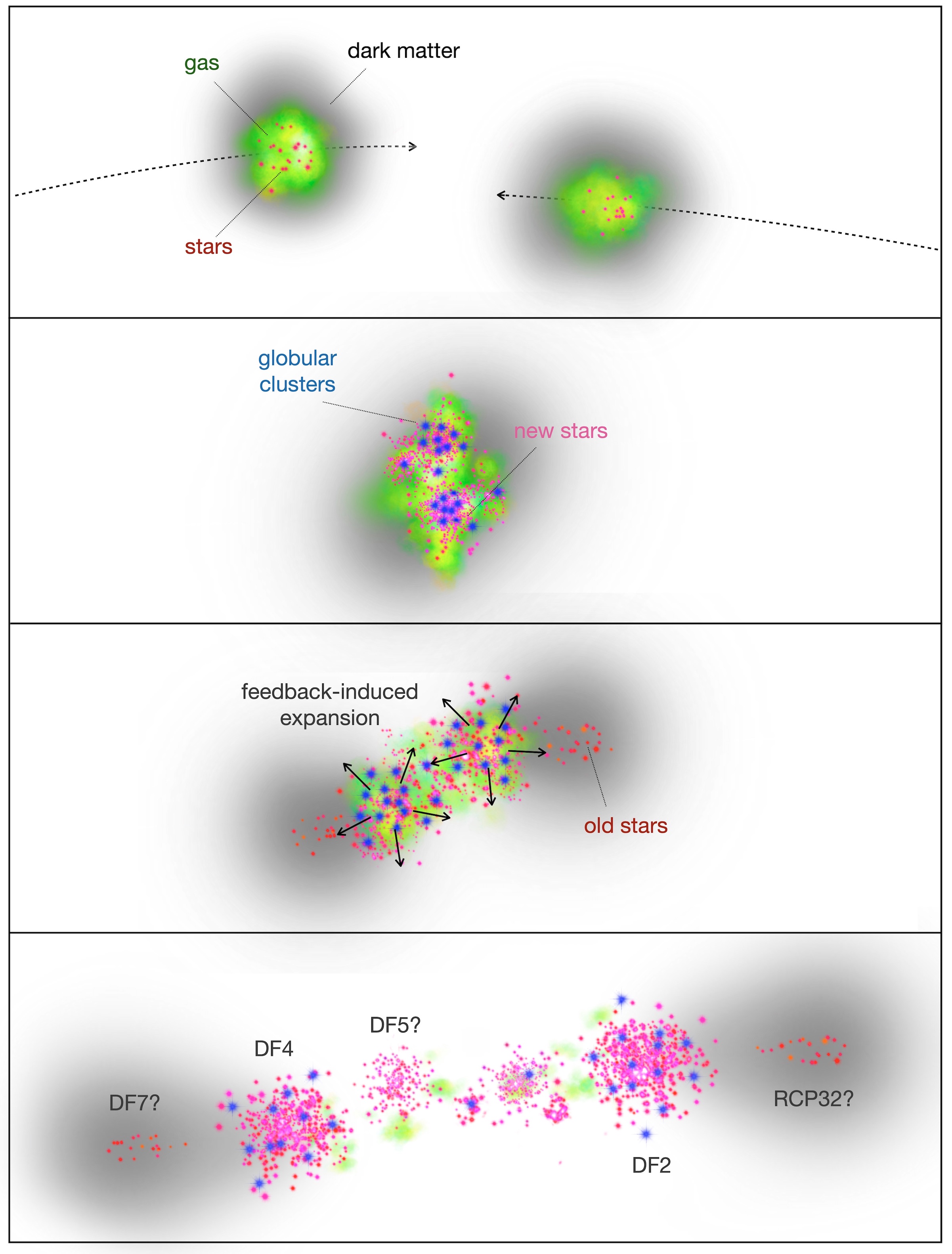}
  \end{center}
    \vspace{-0.3truecm}  
    \caption{\small \textbf{Illustration of the proposed formation scenario of \dft\ and \dff.} 
Two gas rich dwarf galaxies experience a high speed encounter with a small  impact
parameter (top). Following previous studies\cite{silk:19,shin:20,lee:21}
the collisional gas gets stripped and shocked at closest approach, and forms stars at a prodigious rate
with a bias toward massive clumps (second from top). The dark matter and previously formed stars
are tidally distorted but continue ahead of the newly forming galaxies (third from top). Feedback in the absence of
a dark matter halo leads to 
expansion of the newly formed galaxies.\cite{trujillogomez:21} Most of the newly formed stellar mass is in two clumps,
but several lower mass galaxies have also formed in the wake (bottom).
   }
   \label{cartoon.fig}
    \vspace{-12pt}
\end{figure}

\subsection{Faint galaxy sample.}
We make use
of a recently compiled catalog of low surface brightness objects in the NGC1052 field,\cite{roman:21} augmented
by a catalog of
all brighter galaxies with redshifts in the range $1000$\,\kms\,$<cz<$\,2000\,\kms\ that is provided in the
same study. Ref 23 makes use of the publicly available DECaLS dataset.\cite{decals}
The galaxies were initially identified with a combination of automated techniques and visual inspection, with
the majority coming from visual inspection. Their structural parameters were measured with GALFIT.\cite{galfit}
We caution that the DECaLS dataset suffers from sky subtraction errors around low surface brightness galaxies, and
that this may bias the size measurements. The main point of Fig.\ \ref{sizes.fig}
is a relative comparison of the sizes of galaxies on and off the trail and this should be more robust than the absolute size measurements.

\subsection{Velocity dispersion of the NGC1052 group.}  We use the latest compilation of radial velocities in the
NGC1052 field\cite{roman:21} for an updated value of the velocity dispersion of the group. Table 2 of ref 23 contains
30 galaxies with redshifts $cz<2000$\,\kms. Two were removed: DF2, as it is almost certainly not bound to the
group, and LEDA4014647. LEDA4014647 was assigned a radial velocity of $1680\pm 60$ in earlier SDSS releases
(DR3) but its redshift was later revised to $z=0.7$ (DR16). A visual inspection of the SDSS spectrum
shows no clear features. Using the biweight estimator\cite{beers:90} we find a central velocity for the remaining
28 galaxies of $\langle cz \rangle = 1435 \pm 20$\,\kms\ and a line of sight velocity dispersion of $\sigma = 115\pm 15$\,\kms.

\subsection{The Hough transform.}

We use the Hough transform to look for linear features in the galaxy distribution, a standard method for detecting lines in images.\cite{dudahart:72}
The transform provides the number of galaxies along all possible
directions, characterized by an angle and a distance from the center.
A width and maximum linear extent have to be chosen; we use $\pm 30$\,kpc ($\pm 5.2'$) for the
width and  $<400$\,kpc ($69'$) for the linear extent. 
Although the exact number of galaxies that the Hough transform associates with
the linear feature depends on the precise limits that are chosen, the qualitative results are not sensitive to them.
In Fig.\ \ref{spatial.fig}b the orientation of the line is on
the vertical axis and the offset with respect to NGC1052 on the horizontal axis.

\subsection{Statistical significance of the trail.}

We use simulations to assess the probability that the alignment of the 11 galaxies arose by chance. We
generate $N=1000$ realizations of the $(x,y)$ pairs by maintaining the
angular distance from NGC1052 for each pair and randomizing the angle. This procedure ensures
that the density profile of the sample is maintained for all realizations.
We then create Hough transforms for all realizations and
determine how often the strongest linear feature contains $\geq 11$ galaxies.
We find that the probability of a chance
alignment of $\geq11$ galaxies is $3$\,\%. 

This calculation assumes that galaxies are oriented randomly with respect to NGC1052, and does not
take into account anisotropy associated with the filamentary structure of the cosmic web.\cite{bailin:08,tempel:15}
Galaxy groups are generally not spherical but have an average projected axis ratio of 0.77.\cite{wang:08}
We examined the large scale structure in the NGC1052 field using a recently compiled catalog of galaxies in
this general area.\cite{roman:21}  Selecting all low surface brightness galaxies that were identified in that
study plus all bright galaxies with $cz<2000$\,\kms\ gives a sample of 72 probable group members. Their
distribution is shown in Fig.\ \ref{contour.fig}. The smooth density field was
calculated with the non-parametric kernel density estimator.\cite{silverman:86}
There is no evidence for large scale structure associated with the trail. In fact, there are no galaxies at all in the
trail direction in the outskirts of the group, and the overall orientation of the group is perpendicular to the trail.
The assumption of isotropy is therefore slightly conservative, in the sense that  more
galaxies will be scattered toward the line than away from it.

\begin{figure}[htb]
  \begin{center}
  \includegraphics[width=1.0\linewidth]{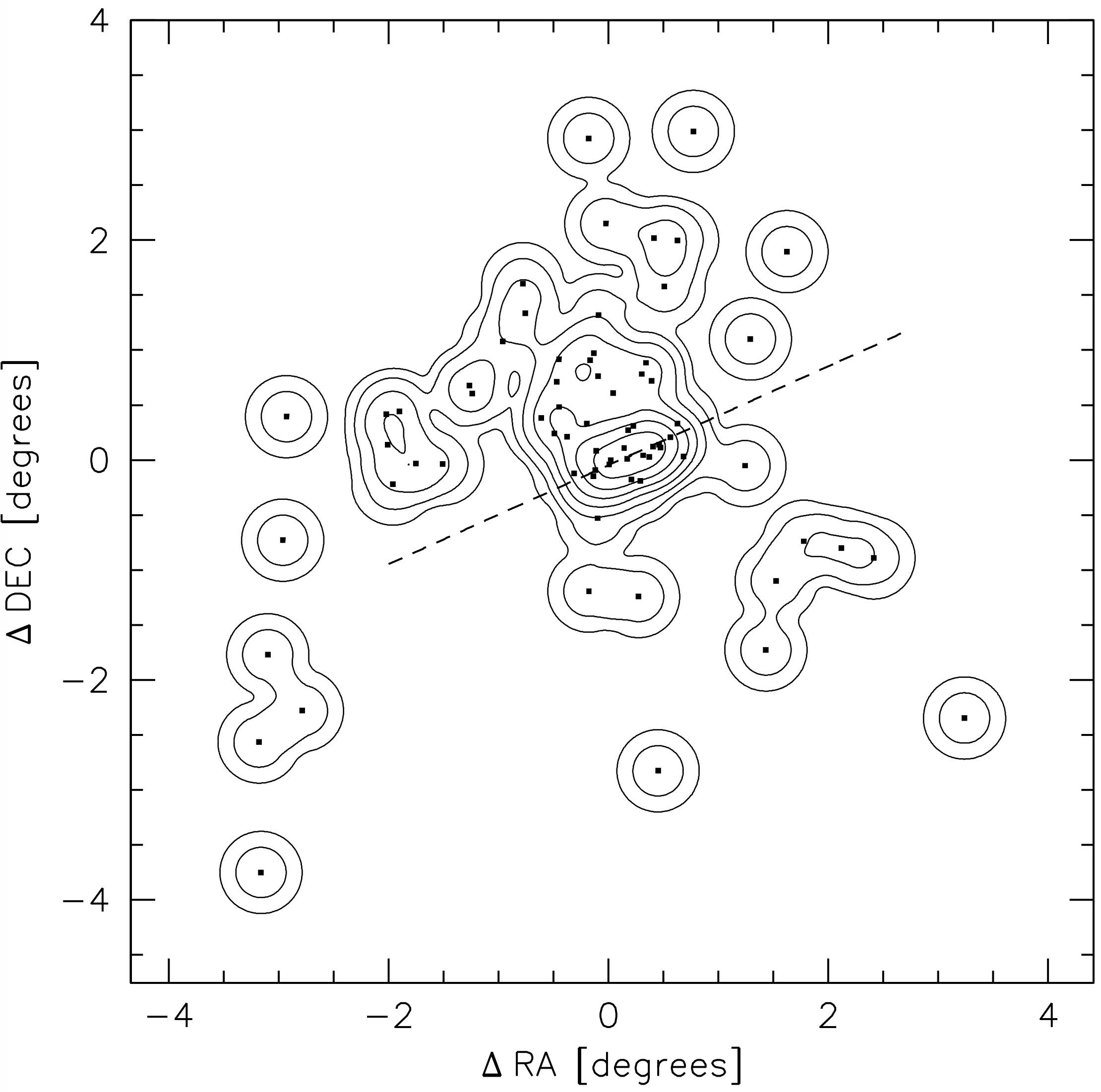}
  \end{center}
    \vspace{-0.5truecm}  
    \caption{\small \textbf{Morphology of the NGC1052 group.} 
Distribution of 72 probable group members from a recent compilation of galaxies in the NGC1052 field.\cite{roman:21}
Contours were derived with the non-parametric kernel density method.\cite{silverman:86} The dashed line indicates the trail.
There is no evidence that the trail is in the general direction of large scale structure in this field.
   }
   \label{contour.fig}
    \vspace{-12pt}
\end{figure}

Finally, we note that the probability that there is a chance alignment
{\em and}  that it is a coincidence that both DF2 and DF4 are part of it is very low. This joint probability can be 
calculated directly for the isotropic case:
of the 31 simulations that have $\geq 11$ aligned galaxies only 6 have both \dft\ and \dff\ 
in the sample, corresponding to a combined probability of the observed arrangement of 0.6\,\%.

\subsection{A twelfth low surface brightness dwarf galaxy on the trail.}

Visual inspection of the DECaLS imaging\cite{decals} readily shows that there is a fairly prominent
12$^{\rm th}$ galaxy that is part of the apparent trail.
The object is SDSS\,J024007.01$-$081344.4; it was previously pointed out as a likely
low luminosity group member with a central star cluster.\cite{trujillo:21}  It is not in the objective catalog that we use for the
main analysis.\cite{roman:21} This may be because of its redshift in the SDSS database (it is erroneously
listed as a $z=0.933$ active galactic nucleus) or because the light from the central cluster moved the object outside of the size and surface brightness
criteria. We refer to the galaxy as DF9 as that was the catalog number in our initial Dragonfly catalog.\cite{cohen:18}
We do not use the galaxy in the objective analysis but
we show its DECaLS image in Fig.\ \ref{cutouts.fig}.
For convenience we provide the coordinates of all trail galaxies in Table 1.

\begin{figure}[htb]
  \begin{center}
  \includegraphics[width=1.0\linewidth]{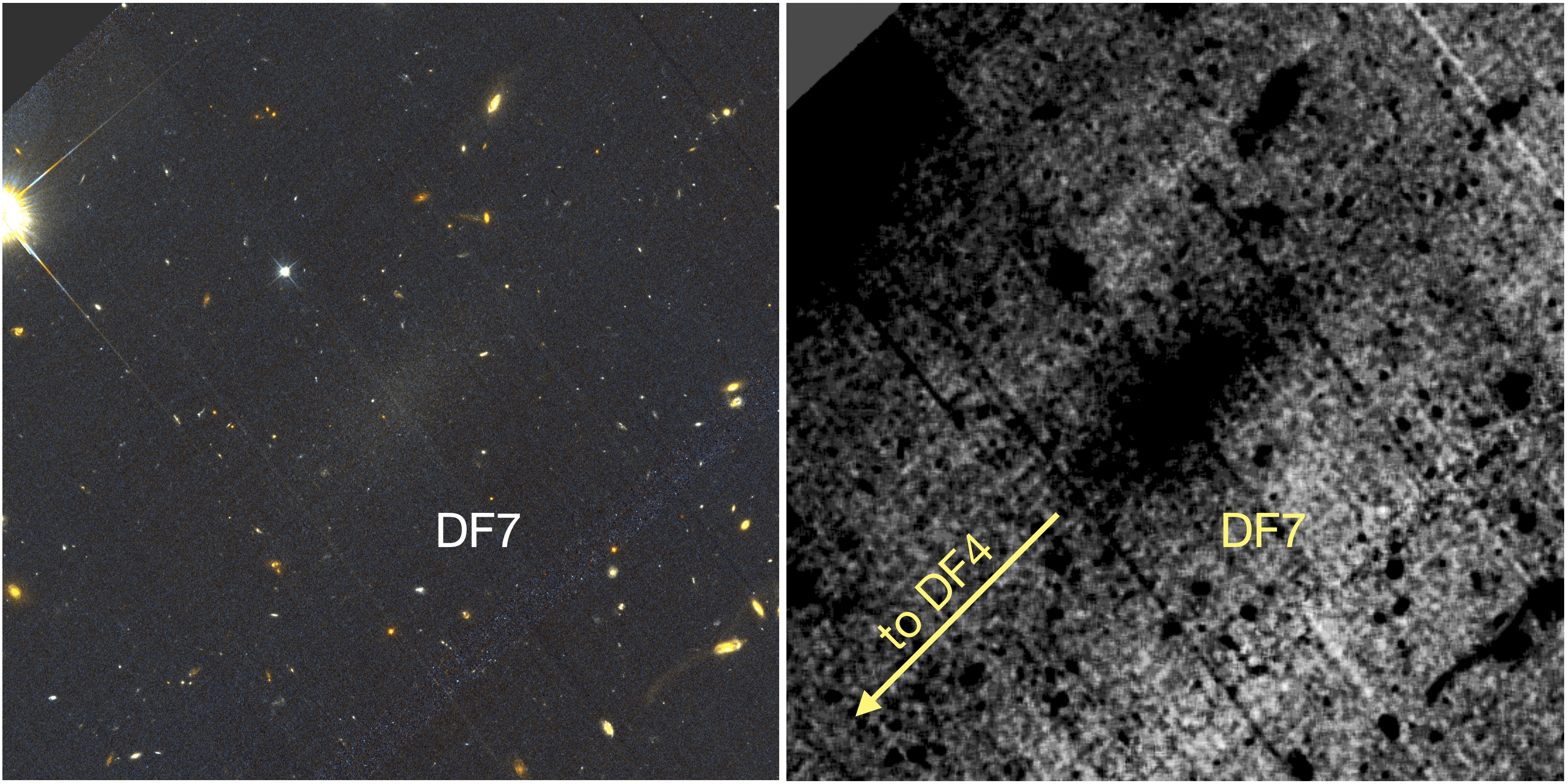}
  \end{center}
    \vspace{-0.5truecm}  
    \caption{\small \textbf{A candidate dark matter-dominated galaxy.} 
HST/ACS images\cite{cohen:18} of DF7, which  is located at the Western end of the galaxy
trail, beyond \dff. At left is a color image generated from the F606W and F814W
data. At right is a median-filtered version of the F606W image, with the arrow depicting the direction toward
DF4.  DF7 is a candidate for
the dark matter-dominated remnant of one of the original galaxies,\cite{silk:19} given its location
and its elongation in the direction of DF4.
   }
   \label{df7.fig}
    \vspace{-12pt}
\end{figure}

\begin{small}
\begin{table}
\centering
 \begin{tabular}{c c c} 
 \hline
 Id & RA [deg] & DEC [deg] \\ 
 \hline\hline
RCP\,32 & $40.6202$  &  $-8.3768$  \\
NGC\,1052--DF2 & $40.4451$  & $-8.4028$   \\
RCP\,28 &  $40.4215$ & $-8.3475$  \\
RCP\,26 &  $40.2897$ &  $-8.2968$ \\
RCP\,21 &  $40.1200$ & $-8.2434$   \\
{\em NGC\,1052--DF9} & $40.0292$  & $-8.2290$  \\
RCP\,17 & $39.9696$  & $-8.2121$  \\
TA\,21--12000 & $39.9139$  &  $-8.2285$ \\
NGC\,1052--DF4 & $39.8128$  &  $-8.1160$ \\
NGC\,1052--DF5 &  $39.8028$ &  $-8.1408$ \\
LEDA\,4014647 & $39.7021$  &  $-8.0494$ \\
NGC\,1052--DF7 &  $39.6241$ &  $-7.9257$\\
 \hline
 \end{tabular}
\caption{Coordinates (J2000) of the trail galaxies.}
\end{table}
\end{small}

\subsection{Hubble Space Telescope imaging of the candidate dark galaxy DF7.}
DF7 is at one of the leading edges of the trail, ``ahead" of DF4. The galaxy was observed with HST/ACS as part of an exploratory survey
of Dragonfly-identified low surface brightness galaxies in several groups.\cite{cohen:18} 
The observations constituted two orbits, one orbit in F606W and one orbit in F814W. 
In Fig.\ \ref{df7.fig} we show the HST imaging at two different contrast levels. 
The galaxy is elongated and appears distorted, with the elongation in the direction of  DF4.  DF7's apparent distortion, combined with its
location at the leading edge of the trail, lead us to speculate that the galaxy is the highly dark matter-dominated remnant of
one of the two progenitor galaxies.
We note that DF7 may be largely disrupted in this interpretation:
the observed\cite{cohen:18} axis ratio is $b/a=0.42$,
but given the extreme foreshortening of the geometry the intrinsic axis ratio could be
a stream-like $\sim 1:20$. 

\subsection{Other proposed scenarios.}
The joint formation of DF2 and DF4 in a single bullet dwarf event explains their lack of dark matter, large sizes, luminous and large globular clusters,
striking similarity, large distance between them, large radial velocity difference, and the presence of a trail of other galaxies on the DF2 -- DF4 axis.
Here we briefly discuss other scenarios that have been proposed to explain the properties of DF2 and DF4.

Initially follow-up studies focused on possible errors in the measurements, either in the masses\cite{martin:18} or in the distances of the galaxies.\cite{trujillo:19,monelli:19}
However, with four independent velocity dispersion measurements\cite{dokkum:18c,dokkum:19df4,danieli:19,emsellem:19} (three for DF2 and one for DF4) and
TRGB distances from extremely deep HST data\cite{danieli:20,shen:21b}  these issues have now largely been settled.

Most astrophysical explanations center on the absence of dark matter only, and invoke some form of extreme tidal interaction (with NGC1052 or other galaxies) to strip the dark matter (along with a large fraction of the initial stellar population).\cite{ogiya:18,maccio:21,jackson:21,moreno:22}  These models do not explain the low metallicity of the galaxies, why there are two nearly identical objects in the same group, the newly discovered trail,
or  their overluminous and too-large globular clusters. The globular clusters,
which have the same age (within the errors) as the diffuse light,\cite{fensch:19} 
show that the galaxies were {\em formed} in an unusual way and did not merely {\em evolve} in an unusual way. 
Besides the bullet scenario, the only model that explains the globular clusters is a study of star formation in galaxies that are in the tails of the scatter in the halo mass -- stellar mass relation.\cite{trujillogomez:21,trujillogomez:21b}
This model has ad hoc initial conditions and does not account for the presence of two near-identical galaxies, but the key aspects of it (the formation of luminous globular clusters in a compact configuration and the subsequent puffing up of the galaxies due to feedback) likely apply to the collision products in the bullet scenario (see main text).

It has recently been suggested that DF2 and DF4 are entirely unrelated, with DF4 being stripped of its dark matter by NGC1035, which is near it in projection, and 
DF2 a face-on disk galaxy with a normal dark matter content.\cite{montes:20,montes:21} The association of DF4 with NGC1035 is not seen in all datasets,\cite{keim:21} and there is no compelling evidence that DF2 is a disk.\cite{dokkum:18}  Furthermore, the globular clusters and the trail remain unexplained, and
there is the question of the likelihood that DF2 and DF4 have entirely different explanations but coindidentally share several otherwise-unique properties.

\subsection{Data availability.}
The HST data for DF7 are available in the Mikulski Archive for Space Telescopes (MAST;
http://archive.stci.edu), under program ID 14644. The Legacy Survey data shown in Fig.\ \ref{cutouts.fig}
are available at https://www.legacysurvey.org/. All other data
that support the findings of this study are available in published studies that are referenced in the text.

\subsection{Code availability.}
We have made use of standard data analysis tools in the Python
environment.

\end{methods}



\vspace{3pt}
\noindent\rule{\linewidth}{0.4pt}
\vspace{3pt}





\begin{addendum}
 \item[Acknowledgements]
A.J.R.\ was supported by National Science Foundation grant
AST-1616710 and as a Research Corporation for Science
Advancement Cottrell Scholar. 
J.M.D.K.\ acknowledges funding from the German Research Foundation (DFG) in the form of an Emmy Noether Research Group grant no.\ KR4801/1-1.\ J.M.D.K.\ and S.T.-G.\ acknowledge funding from the European Research Council (ERC) under the European Union's Horizon 2020 research and innovation programme via the ERC Starting Grant MUSTANG (grant agreement no.\ 714907).
Support from Space Telescope
Science Institute grants HST-GO-14644, HST-GO-15695, and
HST-GO-15851 is gratefully acknowledged. Celeste van Dokkum
created the illustration in Fig.\ \ref{cartoon.fig}.
 \item[Author Contributions] P.v.D.\ led the analysis and wrote the manuscript.  
Z.S.\ performed the relative distance measurement (published in Shen et al.\ 2021)
that is at the basis of the study. M.K.\ created Fig.\ \ref{cutouts.fig} and measured structural parameters
for DF5. S.T.-G.\ created an early version of Fig.\ 1b.  All authors commented
on the manuscript and aided in the interpretation.
 \item[Author Information] The authors declare that they have no
   competing financial interests. Correspondence and requests for
   materials should be addressed to P.v.D.~(email:
   pieter.vandokkum@yale.edu).

\end{addendum}



\appendix

\end{document}